# Intrinsic circular polarization in centrosymmetric stacks of transition-metal dichalcogenides

Qihang Liu[*], Xiuwen Zhang, and Alex Zunger[*]

University of Colorado, Boulder, CO 80309, USA

[*]E-mail: qihang.liu85@gmail.com; alex.zunger@gmail.com.

**Abstract**

The circular polarization (CP) that the photoluminescence inherits from the excitation source in $n$ monolayers of transition-metal dichalcogenides $(MX_2)_n$ has been previously explained as a special feature of *odd* values of $n$, where the inversion symmetry is absent. This "valley polarization" effect results from the fact that in the absence of inversion symmetry, charge carriers in different band valleys could be selectively excited by different circular polarized light. Such restriction to non- centrosymmetric systems poses a limitation on the material selection for achieving CP. Although several experiments observed CP in centrosymmetric $MX_2$ systems, e.g., for bilayer $MX_2$, they were dismissed as being due to some extrinsic sample irregularities. Here we show that also for $n$ = even where inversion symmetry is present and valley polarization physics is strictly absent, such intrinsic selectivity in CP is to be expected on the basis of fundamental spin-orbit physics. First-principles calculations of CP predict significant polarization for $n$ = 2 bilayers: from 69% in $MoS_2$ to 93% in $WS_2$. This realization could broaden the range of materials to be considered as CP sources.

The quest for a system manifesting a single type of circular polarized light – left-handed (LH, or $\sigma_+$) or right-handed (RH, or $\sigma_-$) has been motivated by its promising application in fields such as spin-Hall effect and quantum computation [1-6]. The magnitude of the effect is conveniently measured by circular polarization (CP), i.e., the anisotropy of circular polarized luminescence [7]

$$\rho = \frac{I(\sigma_+) - I(\sigma_-)}{I(\sigma_+) + I(\sigma_-)}, \quad (1)$$

where $I(\sigma_+)$ and $I(\sigma_-)$ denotes the intensity of the LH and RH polarized luminescence, respectively. One approach to the creation of CP is the utilization of *spin-orbit-induced CP* [6] in which spin-orbit coupling (SOC) in low-symmetry bulk solids creates spin splitting (the Dresselhaus [8] and Rashba [9] effects), leading to spin-sensitive absorption and thus polarized luminescence. Another approach to achieving large $\rho$ is the creation of *valley-selective CP* [10]. Here absorption selects the wavefunction character that can couple with the circular polarized light, and SOC is not needed to produce the effect. An example is $n$ = odd monolayers of 2H-stacking transition-metal dichalcogenides $(MX_2)_n$ (with $M$: Mo, W; and $X$: S, Se), having two valleys K and –K with equal energies located at the corners of their hexagonal Brillouin zone. For a single monolayer $n$ = 1, the lowest energy transition is momentum-space direct at the wavevectors K and –K and the wavefunctions of the valence band maximum (VBM) $\psi_v^{(n=1)}(K) = \chi_+$ and $\psi_v^{(n=1)}(-K) = \chi_-$ have pure LH or RH characters, respectively. As a result, $\sigma_+$ excitation can excite the valence electrons at the K valley but no electrons at the –K valley, while $\sigma_-$ can only excite valence electrons at the –K valley but none of those at the K valley. Such valley-contrasting CP leads to valleytronics, a parallel concept of spintronics, with the emergence of interesting phenomena such as valley-Hall effect [11, 12]. However, the limitation to non-centrosymmetric systems both in the spin-orbit induced CP and in the valleytronics-induced CP poses a restriction on material selection for achieving CP.

For a bilayer $MX_2$ ($n$ = 2) the lowest energy transition is momentum-indirect whereas the direct transition at K and –K valley is higher in energy, but the exciton emission from these direct states is still rather strong [14]). More surprisingly, highly selective CP (up to $\rho$ = 80%) has been observed for this direct transition in bilayer $MX_2$ that has inversion symmetry [7, 13-16]. Since this fact goes against the generally accepted expectation that valley-induced CP should be intrinsically absent in centrosymmetric (such as even $n$) materials, various *extrinsic* scenarios were offered to rationalize this unusual observation, such as substrate charging, low-quality sample, and heating effects, etc. [7, 15, 16].

Recently, Gong et al. [17] discussed the magnetoelectric effects of bilayer $MX_2$ using $\boldsymbol{k} \cdot \boldsymbol{p}$ and tight-binding method. They have argued that because the interlayer coupling is weak, bilayers will inherit most of the spin-valley physics of monolayers, named "spin-layer locking". In this Letter, we focus instead on spin-orbit physics as local effect, building on the recent realization [18] that SOC-induced spin polarization can exist not only when inversion symmetry is absent, but also in systems where inversion symmetry is present (i.e., globally centrosymmetric systems) while its individual sectors (e.g., monolayers) lack inversion. We show that such "hidden spin polarization" can lead to CP of the emission from the direct K/–K states for $n$ = even values of $(MX_2)_n$. This is illustrated in Fig. 1 showing our first-principles calculated $\rho$ for the emission from the direct band states at K and –K valley as a function of the number of monolayers $n$ in $(MX_2)_n$, reaching asymptotically the bulk value for a large $n$. We see that the CP decreases monotonically with increasing $n$ and the results of a fixed material lie on one curve without odd-even oscillations, in contrast with the expectation based on valley symmetries [10, 15, 16]. Our "hidden spin polarization" mechanism provides a specific physical realization to the general principle of weak interlayer coupling proposed by Gong et al. By recognizing that the spin-orbit physics can induce CP and that this effect is no longer limited to low-symmetry non-centrosymmetric structures, our finding could broaden the range of materials to be considered as spintronic CP sources.

***Local spin polarization in each monolayer within bilayer $MX_2$.*** This intrinsic CP in centrosymmetric systems originating from "hidden spin polarization" can be illustrated for bilayer $n$ = 2 in $(MX_2)_n$, where two inversion-asymmetric individual $MX_2$ layers α and β ("sectors" in general) carry opposite local spin polarization. In bilayer $MX_2$, a monolayer $MX_2$ named β is introduced to form inversion partner of layer α. The corresponding energy bands must be spin-degenerate due to the combination of inversion symmetry and time-reversal symmetry. However, such global $k$-space compensation of spins does not occur in a point-by-point fashion in real space, i.e., on each $MX_2$ layer.

Using density functional theory (DFT) calculation implemented by VASP [19] with projected augmented wave (PAW) pseudopotential [20], we can project the two-fold degenerate wavefunctions with plane-wave expansion on the spin and orbital basis (spherical harmonics) of each atomic site. For K valley of the top valence band (V1),

$$\psi_v^{(n=2)}(K,\uparrow) = \sum_{l,m,i} C_{l,m,i,\uparrow} |l,m,i\rangle \otimes |\uparrow\rangle,$$
$$\psi_v^{(n=2)}(K,\downarrow) = \sum_{l,m,i} C_{l,m,i,\downarrow} |l,m,i\rangle \otimes |\downarrow\rangle, \quad (2)$$

with the module squared expansion coefficient

$$\left|C_{i,l,m,\eta}\right|^2 = \left\langle \psi_v^{(n=2)}(K,\eta) \middle| (s_z \otimes |l,m,i\rangle\langle l,m,i|) \middle| \psi_v^{(n=2)}(K,\eta) \right\rangle \quad (3)$$

where $\psi_v^{(n=2)}(K,\eta)$ is the hole state of bilayer $MX_2$, $|l,m,i\rangle$ is the orbital angular momentum eigenstate centered about the $i$th atomic site, $s_z = \frac{\hbar}{2}\sigma_z$ is the spin operator, and $\eta$ denotes spin. Note that at K and –K valley $s_z$ is a good quantum number with two spin eigenstates $|\uparrow\rangle$ and $|\downarrow\rangle$. By summing $C_{i,l,m,\eta}$ separately for layer α and for layer β (each containing 3 atomic sites in a unit cell), we find that for K valley of V1 in bilayers of MoS$_2$, MoSe$_2$, WS$_2$, and WSe$_2$ the spin polarization localized on layer α, $S = \sum_{l,m,i\in\alpha}(|C_{i,l,m,\uparrow}|^2 - |C_{i,l,m,\downarrow}|^2)$, is 0.83, 0.90, 0.97 and 0.95 (out-of plane, in the unit of $\frac{\hbar}{2}$) respectively, and the local spin polarization on layer β has exactly the same magnitude but opposite direction. Thus, despite the global inversion of the $n = 2$ structure, each individual layer experiences a nonzero local spin polarization $S$ in real space, "hidden" by the presence of the counterpart on the other layer. We will next show that in such even-layer stacks of $(MX_2)_n$ the hidden spin polarization induce CP as an intrinsic part of the fundamental physics even without any sample irregularities.

*Photoluminescence process leading to circular polarization.* To calculate CP of Eq. (1) that involves *emission* experiment, we need a model of what happens to the photo excited carriers before emission. Taking bilayer $MX_2$ as an example, we will next consider the absorption, relaxation and emission process, as illustrated in Fig. 2. The splitting between the top two valence bands V1 and V2 at K and –K is due to SOC-induced spin splitting and the interlayer coupling. Here we consider resonant excitation (the incident photon energy equals to the direct band gap at K or –K), in which only V1 is of interest.

Interband absorption is described by the up arrow indicated as ① in Fig. 2(a). It involves promoting an electron from a valence band eigenstate of Eq. (2) momentum-directly into a conduction band eigenstate, evaluated by the transition matrix element $\mathcal{P}_\pm = \langle \psi_c | p_x \pm i p_y | \psi_v \rangle$, where ***p*** is the momentum operator. By calculating the transition matrix element for circular polarized light, say $\sigma_+$, we find that unlike the case of one monolayer in which absorption is valley contrasting, in bilayer *MX*$_2$ the K and –K valley have the same absorption, consistent with the presence of global inversion symmetry [10]. However, our present DFT calculations show that the two degenerate spin states in Eq. (2) have different non-zero absorption magnitude, leading to a net spin of the excited electrons. To interpret this effect and its impact on CP, we reconsider the valence band eigenstates of the bilayer system in Eq. (2) expanded in terms of single monolayer eigenstates, $\psi_v^{(n=1)}(K) = \chi_+$ and $\psi_v^{(n=1)}(-K) = \chi_-$ having pure LH or RH characters, respectively. We approximate the wavefunction of bilayer *MX*$_2$ at K valley as an expansion in terms of the α layer ($\chi_+^{(\alpha)}$) and –K valley of β layer ($\chi_-^{(\beta)}$) eigenstates, written as

$$\psi_v^{(n=2)}(K,\uparrow) = (C_{\alpha,\uparrow}\chi_+^{(\alpha)} + C_{\beta,\uparrow}\chi_-^{(\beta)})\otimes|\uparrow\rangle,$$
$$\psi_v^{(n=2)}(K,\downarrow) = (C_{\alpha,\downarrow}\chi_+^{(\alpha)} + C_{\beta,\downarrow}\chi_-^{(\beta)})\otimes|\downarrow\rangle, \quad (4)$$

where the module squared coefficient is $|C_{\alpha(\beta),\eta}|^2 = \sum_{l,m,i\in\alpha(\beta)}|C_{i,l,m,\eta}|^2$, and $C_{\alpha,\uparrow} = C_{\beta,\downarrow} = C_{maj}$, $C_{\beta,\uparrow} = C_{\alpha,\downarrow} = C_{min}$ denote the majority (maj) and minority (min) component for each valence spin state. Subscript + (–) denotes pure LH (RH) wavefunction component. The wavefunctions of V1 at –K valley is just related to those at K valley by time reversal

$$\psi_v^{(n=2)}(-K,\uparrow) = (C_{maj}\chi_+^{(\beta)} + C_{min}\chi_-^{(\alpha)})\otimes|\uparrow\rangle,$$
$$\psi_v^{(n=2)}(-K,\downarrow) = (C_{min}\chi_+^{(\beta)} + C_{maj}\chi_-^{(\alpha)})\otimes|\downarrow\rangle. \quad (5)$$

We have tested the approximation of expanding *n* = 2 just by the two *n* = 1 eigenstates by comparing the calculated ratio of transition matrix elements $\mathcal{P}_+(K,\uparrow)/\mathcal{P}_+(K,\downarrow)$ of the corresponding spin states with the initial plane-wave expansion via first-principles calculation, and find it in good agreement with $|C_{maj}|^2/|C_{min}|^2$ from Eq. (4) and (5), indicating the validity of our wavefunction decomposition onto the monolayer basis.

Unlike the case of one monolayer in which $\psi_v^{(n=1)}(\pm K)$ have pure LH or RH components ($\chi_+$ and $\chi_-$), the four valence states of bilayer $MX_2$, described by Eq. (4) and (5), have both $\chi_+$ and $\chi_-$ components. Therefore, for $\sigma_+$ resonant excitation only the electrons with $\chi_+^{(\alpha)}$ character at the K valley and the electrons with $\chi_+^{(\beta)}$ character at the –K valley are excited ($\chi_-$ components have no contribution to the transition matrix element), as shown in Fig. 2 by step ①. Note that for $\psi_v^{(n=2)}(K,\uparrow)$ and $\psi_v^{(n=2)}(-K,\uparrow)$ eigenstates [Fig. 2(a) and (c)], the excited component is the majority term (proportional to $|C_{maj}|^2$) with up-spin $|\uparrow\rangle$, while for $\psi_v^{(n=2)}(K,\downarrow)$ and $\psi_v^{(n=2)}(-K,\downarrow)$ states [Fig. 2(b) and (d)], the excited component is the minority term (proportional to $|C_{min}|^2$) with down-spin $|\downarrow\rangle$. As a result, the excited electrons have a net up-spin that equals to the local spin polarization on α layer at K valley (or β layer at –K) derived from Eq. (4) and (5), $S = \frac{|C_{maj}|^2 - |C_{min}|^2}{|C_{maj}|^2 + |C_{min}|^2}$, creating simultaneously equivalent holes with the same net spin.

Following excitation, relaxation mechanisms might redistribute the photogenerated LH and RH holes, thereby reducing the polarization anisotropy $\rho$. The observed $\rho$ for monolayer $MoS_2$ has been as large as 100% [7], suggesting ineffective mixture of photogenerated LH and RH holes caused by intervalley relaxation and thus excellent retention of the valley-contrasting circular absorption. In bilayer $MX_2$, the dominated relaxation route considered here is illustrated by the green arrow labeled step ② in Fig. 2(a). For the $\sigma_+$ excitation, all of the promoted electrons have $\chi_+$ character, leaving the excited state with a different LH/RH ratio compared with the ground state. Consequently, the LH and RH components tends to redistribute within the same valence eigenstate to retain the ground state ($\chi_- \to \chi_+$). Such intrastate relaxation could happen much faster than the electron-hole recombination because of the imbalance excitation between $\chi_+$ and $\chi_-$ components and the lack of effective barrier to suppress the spin-conserved relaxation. Three kinds of other relaxation channels that could mix LH and RH components and thus affect $\rho$ are also considered. We assume the relaxation time with spin flip much slower, while the relaxation time with spin conserving much faster, than the electron-hole recombination time, with the details listed in Supplementary Materials Section B [21].

As indicated by red and blue curled arrows (step ③), the radiative recombination fills the holes with $\chi_+$ and $\chi_-$ characters of valence band, leading to $\sigma_+$ and $\sigma_-$ luminescence, respectively. In Fig. 2(a) and (c), the excited $\chi_+$ component is majority, so the total emission is larger than that in Fig. 2(b) and (d), in which excited $\chi_+$ component is minority. Note that the magnitude of $\sigma_+$ and $\sigma_-$ emission in each state reflects the comparison of $\chi_+$ and $\chi_-$ components. The Supplementary Materials Section A considers the matrix elements (e.g., electric dipole) of interband absorption under resonant excitation [Eq. (S1)-(S6)]. After relaxation and emission processes take place [Eq. (S7)-(S11)], the expression of the total CP defined by Eq. (1) emerges [21],

$$\rho = \left(\frac{|C_{maj}|^2 - |C_{min}|^2}{|C_{maj}|^2 + |C_{min}|^2}\right)^2. \tag{6}$$

Similarly, we can get exactly opposite $\rho$ by using $\sigma_-$ resonant excitation (see Fig. S1 [21]). From Eq. (6) we find that the polarization anisotropy $\rho$ is closely related to *S*: there is no intrinsic CP unless there is local spin on each *MX*$_2$ layer. Generally, if a centrosymmetric material has local spin polarization on each inversion-asymmetric sector, we can find non-zero CP following excitation of circular polarized light, as all the *MX*$_2$ bilayers shown in Fig. 1.

The calculated CP [Eq. (6)] refers to the direct transition at K and –K valley no matter this transition is the lowest energy transition (as in monolayer) or a higher energy transition accessed by resonant excitation (as in bilayer). In the latter case there is a non-radiative electron relaxation from the absorbing, excited conduction band edge at K valley to the lower conduction band minimum (CBM). This process naturally causes a reduction of intensity of the photoluminescence from the excited conduction band, which may in turn contribute to some uncertainty in its measurement (and thus could be responsible for some of the deviation with theory). Recent experiments [14, 22] show, however, that although the photoluminescence of the direct band gap energy at K (–K) valley in bilayer has a much lower intensity than that in the monolayer case (less than 10%), the intensity is still sufficient to measure the CP. Note that the central quantity here – the ratio between emission intensities at different polarizations [Eq. (1)] – need not to be affected by the fundamental transition being direct or indirect. Indeed our calculated

CP (Fig. 1) does not manifest any non-monotonicity with layer thickness $n$ expected of the intensity of the lowest energy transition due to its direct/indirect nature.

***Progression of CP from n = 1 to bulk $MX_2$.*** From the above discussion we note that the CP of bilayer $MX_2$ is directly related to the spin polarization localized on each layer, rather than the global inversion symmetry. This dependence provides us the route to evaluate CP for all $(MX_2)_n$ stacks by straightforwardly calculating the ratio of LH/RH components of the valence band wavefunctions expanded by monolayer basis. For 2H-$MX_2$ stacking pattern, the K (–K) valley of each monolayer in $(MX_2)_n$ has pure LH or RH component alternatively, depending on the odd-even parity of the layer index. When $n > 1$, The band edges of $(MX_2)_n$ at K and –K valley, no matter degenerate or non-degenerate, have no longer a certain helicity, but a mixture of LH and RH characters from different layers, leading to the reduction of CP away from the case of monolayer. If the orbital part of (one of) the valence band wavefunction at K or –K valley of $(MX_2)_n$ is written as

$$\psi_v = C_1\chi_+^{(1)} + C_2\chi_-^{(2)} + C_3\chi_+^{(3)} + \cdots + C_n\chi_{+/-}^{(n)}, \tag{7}$$

where $C_i$ $(i = 1, 2, \ldots, n)$ is the coefficient of LH or RH component from the $i$th layer, we can thus obtain the corresponding theoretical circular polarization by resonant excitation

$$\rho = \left(\frac{\sum_n (-1)^{n-1}|c_n|^2}{\sum_n |c_n|^2}\right)^2. \tag{8}$$

Note that although the spin part is not included to calculate the CP, it plays a crucial role for the spin-conserving intrastate relaxation and the prevention of spin-flip relaxation. The CP of four $(MX_2)_n$ materials as a function of the layer number $n$ is shown in Fig. 1. Contrasting with the conventional understanding that the CP should oscillate with odd-even layers due to the absence or presence of inversion symmetry, we found a decreasing polarization ratio with $n$. The occurrence of a regular decrease of the CP in Fig. 1 reflects the reduction in the spin polarization piece localized on each layer as the number of layers $n$ increases, a decrease caused by the larger mixture of LH and RH components. In addition, all of $(WX_2)_n$ stacks have larger ρ than that of $(MoX_2)_n$ due to the larger SOC of W atom. All the CP curves reveal an asymptotic behavior to the bulk value. This is because starting from $n = 1$ in which the valence states have pure LH or RH component, the ratio between LH/RH tends to saturate as $n$ increases.

Such trend is confirmed by a recent experiment on few-layer 2H-MoS$_2$ [22], which also exhibits a decreasing dependence on layer number with saturation, indicating a good agreement with our prediction except for the starting point of CP in monolayer (about 58%, possibly because of the sample quality). This agreement also suggested that the formation of exciton states, not considered in this work, might not change significantly the calculated CP. Some other experiments also detected nonzero polarized luminescence $\rho_{exp}$ of $MX_2$ bilayers, but smaller than our calculated theoretical limit (e.g., 14% lower for bilayer WS$_2$ [14]). We list the possible reasons in the Supplementary Materials Section C [21]. Furthermore, because of the local spin on each $MX_2$ layer, the CP range from 37% (MoS$_2$) to 83% (WS$_2$) even for 2H-stacking *bulk* $MX_2$ crystal. Recently, the "hidden spin polarization" effect in bulk WSe$_2$ has been observed experimentally by spin- and angle-resolved photoemission spectroscopy (SR-ARPES) [23], revealing large layer-dependent local spin polarization. Therefore, we expect the intrinsic CP in centrosymmetric bulk $MX_2$ to be realized by upcoming measurements.

***Dependence of ρ on the interlayer distance and material design for larger CP***: The dependence of $\rho$ on the interlayer distance for different bilayer $MX_2$ compounds is shown in Fig. 3. The curves clearly exhibit that the intrinsic CP is enhanced as the interlayer separation increases, which could be achieved by tensile strain along stacking direction or within the two-dimensional (2D) plane [24]. At the equilibrium separation (experimental values indicated by solid symbols), our calculated CP is 69% for MoS$_2$, 81% for MoSe$_2$, 90% for WSe$_2$, and 93% for WS$_2$ [25]. We further note that for the same interlayer distance, the CP of W$X_2$ is larger than that of Mo$X_2$. Furthermore, when the interlayer distance of the $MX_2$ bilayer exceeds 4 Å, the coupling between α and β layer becomes negligible, implying perfect local spin polarization. As a result, the CP approaches monolayer limit $\rho = 1$ when the interlayer distance is large enough.

Using the understanding of hidden spin polarization, we design a heterostructure with larger CP by intercalating 2D bilayer BN as inert medium into bilayer MoSe$_2$ (see Supplementary Materials Section D [21]). Such sandwiched structures, having both optimized polarization anisotropy $\rho \approx 1$ and large photoluminescence intensity due to their direct bandgap, could be good platforms to realize intrinsic CP in a centrosymmetric system by the current synthesis technology [26, 27].

In summary, by using first-principles calculations, we demystify the occurrence of intrinsic CP, accessed by direct interband transition at K and –K valley, in centrosymmetric layer stacks made of individually non-centrosymmetric layers, such as $n$ = even $(MX_2)_n$. The intrinsic CP decreases monotonically with increasing $n$, in sharp contrast with the conventional expectation of odd-even oscillations based on valley symmetries. Such polarization anisotropy results from hidden spin polarization that is localized on each $MX_2$ layer. Our finding is expected to broaden the material selection that is currently limited to non-centrosymmetric systems for achieving CP and related phenomena, and provide new possibility for the manipulation of spin in the field of spintronics and optoelectronics.

**Acknowledgements**

We are grateful for the good discussions with Yu Ye from UC Berkeley, Yang Song from Rochester University, Xiaobo Yin from University of Colorado and helpful codes for the calculation of transition matrix elements from Kanber Lam, Northwestern University. This work was supported by NSF Grant No. DMREF-13-34170. This work used the Extreme Science and Engineering Discovery Environment (XSEDE), which is supported by National Science Foundation grant number ACI-1053575.

**Reference**

[1] A. Crepaldi *et al.*, Phys. Rev. B **89**, 125408 (2014).
[2] T. D. Nguyen, E. Ehrenfreund, and Z. V. Vardeny, Science **337**, 204 (2012).
[3] V. L. Korenev, I. A. Akimov, S. V. Zaitsev, V. F. Sapega, L. Langer, D. R. Yakovlev, Y. A. Danilov, and M. Bayer, Nat. Commun. **3**, 959 (2012).
[4] M. Ghali, K. Ohtani, Y. Ohno, and H. Ohno, Nat. Commun. **3**, 661 (2012).
[5] J. Wunderlich, B. Kaestner, J. Sinova, and T. Jungwirth, Phys. Rev. Lett. **94**, 047204 (2005).
[6] F. Meier and B. P. Zakharchenya, *Optical Orientation* (Elsevier Science Publishers B.V., 1984).
[7] K. F. Mak, K. He, J. Shan, and T. F. Heinz, Nat. Nanotech. **7**, 494 (2012).
[8] G. Dresselhaus, Phys. Rev. **100**, 580 (1955).
[9] E. I. Rashba, Soviet Physics-Solid State **2**, 1109 (1960).
[10] W. Yao, D. Xiao, and Q. Niu, Phys. Rev. B **77**, 235406 (2008).
[11] D. Xiao, G.-B. Liu, W. Feng, X. Xu, and W. Yao, Phys. Rev. Lett. **108**, 196802 (2012).
[12] X. Xu, W. Yao, D. Xiao, and T. F. Heinz, Nat. Phys. **10**, 343 (2014).

[13] A. M. Jones, H. Yu, J. S. Ross, P. Klement, N. J. Ghimire, J. Yan, D. G. Mandrus, W. Yao, and X. Xu, Nat. Phys. **10**, 130 (2014).
[14] H. Z. Bairen Zhu, Junfeng Dai, Zhirui Gong, Xiaodong Cui, Proc. Natl. Acad. Sci. **111**, 11606 (2014).
[15] S. Wu *et al.*, Nat. Phys. **9**, 149 (2013).
[16] H. Zeng, J. Dai, W. Yao, D. Xiao, and X. Cui, Nat. Nanotech. **7**, 490 (2012).
[17] Z. Gong, G.-B. Liu, H. Yu, D. Xiao, X. Cui, X. Xu, and W. Yao, Nat. Commun. **4** 2053 (2013).
[18] X. Zhang, Q. Liu, J.-W. Luo, A. J. Freeman, and A. Zunger, Nat. Phys. **10**, 387 (2014).
[19] G. Kresse and J. Furthmüller, Comp. Mater. Sci. **6**, 15 (1996).
[20] G. Kresse and D. Joubert, Phys. Rev. B **59**, 1758 (1999).
[21] See Supplementary Material at [URL] for the derivation of Eq. (6), three kinds of other relaxation channels, possible reasons for the deviation between theory and experiment, material design for larger CP and supplementary figures.
[22] R. Suzuki *et al.*, Nat. Nanotech. **9**, 611 (2014).
[23] J. M. Riley *et al.*, Nat. Phys. **10,** 835 (2014).
[24] L. Dong, A. M. Dongare, R. R. Namburu, T. P. O'Regan, and M. Dubey, Appl. Phys. Lett. **104,** 043715 (2014).
[25] We noted that the calculated circular polarization does not change if the $MX_2$ layers are slightly shifted or rotated relative to each other.
[26] G.-H. Lee *et al.*, ACS Nano **7**, 7931 (2013).
[27] C. L. Heideman, S. Tepfer, Q. Lin, R. Rostek, P. Zschack, M. D. Anderson, I. M. Anderson, and D. C. Johnson, J. Am. Chem. Soc. **135**, 11055 (2013).

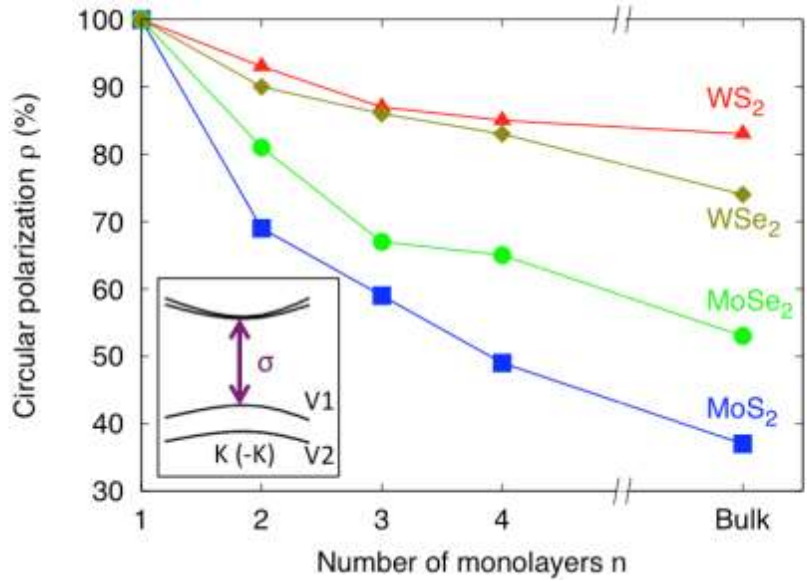

Fig. 1: Calculated circular polarization $\rho$ for different $(MX_2)_n$ materials as a function of the number of monolayers $n$. $\rho$ corresponds to the direct states at K and –K valley no matter they correspond to the lowest-energy excitation (as in $n = 1$) or higher-energy resonant excitation (as in $n > 1$) of circular polarized light. The insert shows a schematic band structure of bilayer $MX_2$. The results by $\sigma_+$ and $\sigma_-$ excitation have the same magnitude but opposite sign.

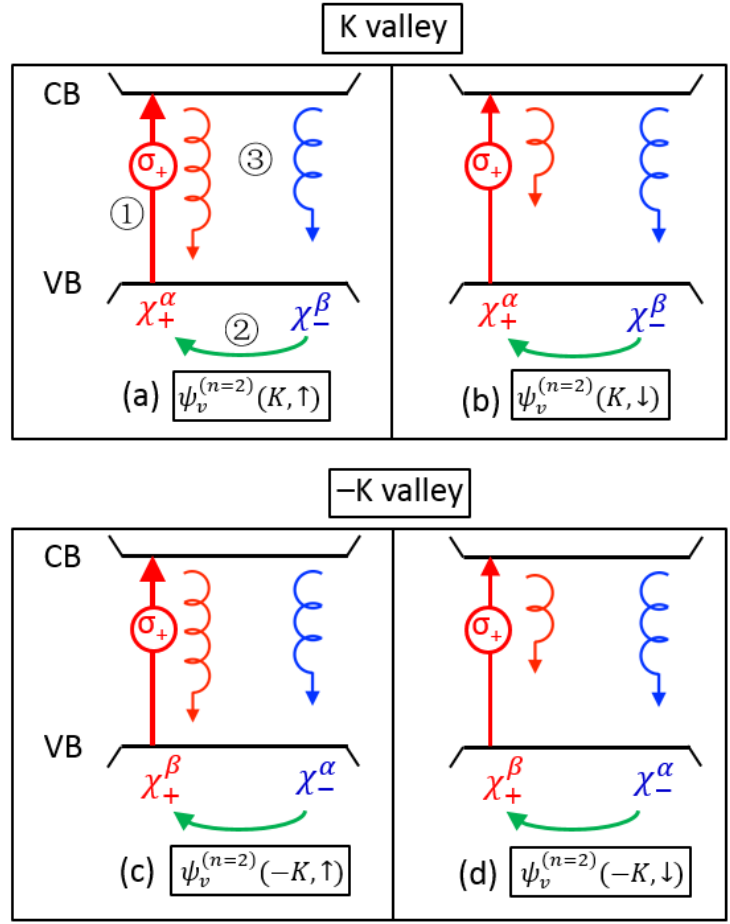

Fig. 2: Schematic diagram of photoluminescence process at K and –K valley in bilayer *MX*$_2$. Four degenerate valence band (VB) states described in Eq. (4) and (5), i.e., (a) $\psi_v^{(n=2)}(K,\uparrow)$, (b) $\psi_v^{(n=2)}(K,\downarrow)$, (c) $\psi_v^{(n=2)}(-K,\uparrow)$, and (d) $\psi_v^{(n=2)}(-K,\downarrow)$, are shown with $\chi_+$ (red) and $\chi_-$ (blue) denoting LH and RH orbitals of the valence states, respectively. Taking panel (a) as an example, three steps in photoluminescence are indicated. ① Absorption by a resonant $\sigma_+$ excitation. The red arrow indicates that the excitation promotes electrons with $\chi_+$ character into conduction band (CB). ② Intrastate relaxation denoted by green arrows. ③ Radiative recombination filling the holes of VB state. The red and blue curled arrows denote $\sigma_+$ and $\sigma_-$ emission, respectively, with the lengths representing the relative intensity of emission regarding $C_{maj} > C_{min}$.

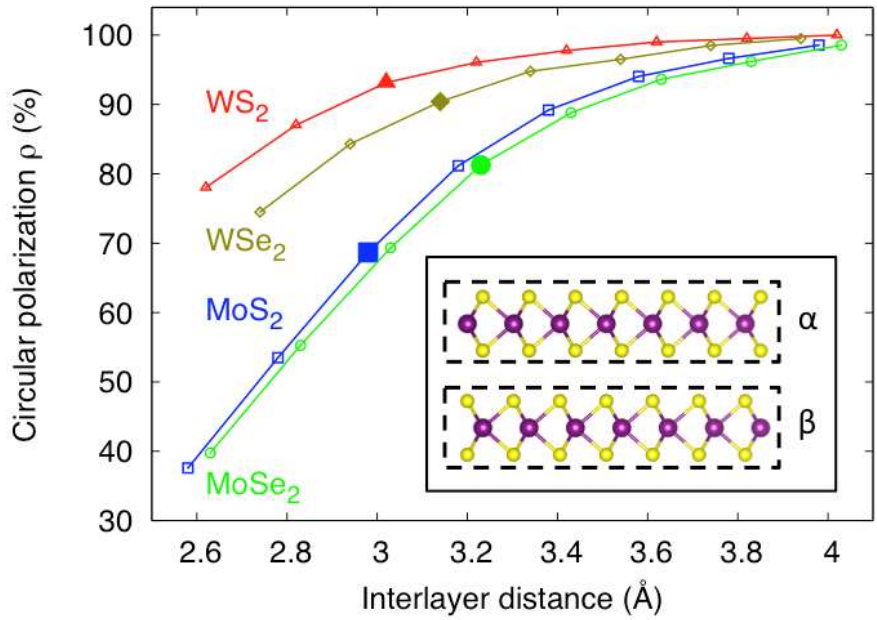

Fig. 3: The circular polarization $\rho$ as a function of the interlayer distance of bilayer $MX_2$ materials. The larger solid symbols denote the experimental interlayer distance. The inset shows the crystal structure of bilayer $MX_2$, with the purple and yellow balls denoting $M$ and $X$ atoms, respectively.

# Supplementary Material
# Intrinsic circular polarization in centrosymmetric stacks of transition-metal dichalcogenides

Qihang Liu*, Xiuwen Zhang, and Alex Zunger*

University of Colorado, Boulder, CO 80309, USA

*E-mail: qihang.liu85@gmail.com; alex.zunger@gmail.com.

### A. Derivation of circular polarization ρ in bilayer MX₂.

We will consider absorption, relaxation and emission processes step by step to derive the circular polarization $\rho$ in bilayer $MX_2$ after a resonant excitation by circular polarized light, say $\sigma_+$.

For monolayer $MX_2$, the conduction band edge (C1) at K and –K valley are almost $d$ states of $M$ atom with magnetic quantum number $m = 0$, and thus have no LH or RH helicity, with the neutral wavefunctions $\psi_c^{(n=1)}(K) = \psi_c^{(n=1)}(K) = \chi_0$. On the other hand, the wavefunctions of the valence band edge (V1) at these wavevectors are $\psi_v^{(n=1)}(K) = \chi_+$ and $\psi_v^{(n=1)}(-K) = \chi_-$, which have pure LH and RH characters, respectively. Therefore, we have the matrix elements of interband transition in the following:

$$\mathcal{P}_+(K) = \langle \chi_0 | p_x + ip_y | \chi_+ \rangle = P. \tag{S1}$$

$$\mathcal{P}_-(K) = \langle \chi_0 | p_x - ip_y | \chi_+ \rangle = 0, \tag{S2}$$

$$\mathcal{P}_+(-K) = \langle \chi_0 | p_x + ip_y | \chi_- \rangle = 0. \tag{S3}$$

$$\mathcal{P}_-(-K) = \langle \chi_0 | p_x - ip_y | \chi_- \rangle = P, \tag{S4}$$

As a result, $\sigma_+$ excitation can excite the valence electrons at the K valley but none of those at the –K valley, while $\sigma_-$ can only excite valence electrons at the –K valley but not at the K valley. We don't need to consider spin selective rule here because C1 states contains both spin-up and spin-down states that can match any spin state of each excitation from V1 when orbital selective rule is fulfilled. Following excitation, however, relaxation mechanisms might redistribute the photogenerated LH and RH holes, thereby reducing the polarization anisotropy $\rho$. We note that in monolayer $MX_2$, only LH holes is generated by $\sigma_+$ excitation. Therefore, the theoretical estimation of $\rho$ in monolayer $MX_2$ is 100%. The experimentally observed $\rho$ for monolayer MoS$_2$ has been as large as 100% [1], suggesting that ignoring the intervalley relaxation, which requires spin-flip energy and large phonon compensation, is a reasonable assumption.

For bilayer $MX_2$, the C1 state at K and –K valley is almost four-fold degenerate with $\chi_0^{(\alpha)} \otimes |\uparrow\rangle$, $\chi_0^{(\alpha)} \otimes |\downarrow\rangle$, $\chi_0^{(\beta)} \otimes |\uparrow\rangle$, and $\chi_0^{(\beta)} \otimes |\downarrow\rangle$, while the V1 state is given in Eq. (4) and (5) in the main text. For $\sigma_+$ resonant excitation at K valley, the transition matrix element is

$$\mathcal{P}_+(K,\uparrow) = \left\langle \psi_c^{(n=2)}(K,\uparrow) \middle| p_x + ip_y \middle| \psi_v^{(n=2)}(K,\uparrow) \right\rangle = C_{maj}P \tag{S5}$$

$$\mathcal{P}_+(K,\downarrow) = \left\langle \psi_c^{(n=2)}(K,\downarrow) \middle| p_x + ip_y \middle| \psi_v^{(n=2)}(K,\downarrow) \right\rangle = C_{min}P \tag{S6}$$

where (S1) and (S3) are taken into account. From Eq. (S5) and (S6) we can see during the absorption process only part of the V1 states that contains $\chi_+$ component is allowed to excite.

After excitation, for state $\psi_v^{(n=2)}(K,\uparrow)$, the number of photogenerated holes is proportional to $|C_{maj}P|^2$. The residual electrons occupying $\chi_-^{(\beta)}$ can relax to empty $\chi_+^{(\alpha)}$ to retain the ratio between LH and RH in the ground state. We are referring to time scales where the ultrafast process of redistribution of the hole state has already taken place but the excited state has not relaxed yet via electron-hole recombination, and then the number of holes with LH or RH components (denoted as "+" and "-") is

$$N_+(K,\uparrow) = \frac{|C_{maj}|^2}{|C_{maj}|^2 + |C_{min}|^2} |C_{maj}P|^2 \tag{S7}$$

$$N_-(K,\uparrow) = \frac{|C_{min}|^2}{|C_{maj}|^2 + |C_{min}|^2} |C_{maj}P|^2 \tag{S8}$$

Similarly, for state $\psi_v(K,\downarrow)$ we have,

$$N_+(K,\downarrow) = \frac{|C_{min}|^2}{|C_{maj}|^2 + |C_{min}|^2} |C_{min}P|^2 \tag{S9}$$

$$N_-(K,\downarrow) = \frac{|C_{maj}|^2}{|C_{maj}|^2 + |C_{min}|^2} |C_{min}P|^2 \tag{S10}$$

The final process is emission, where the radiative recombination fills the holes with $\chi_+$ and $\chi_-$ components of valence band, leading to $\sigma_+$ and $\sigma_-$ luminescence, respectively. According to Eq. (1) and (S5)-(S10), we have circular polarization as shown in Eq. (6)

$$\rho = \frac{I(\sigma_+)-I(\sigma_-)}{I(\sigma_+)+I(\sigma_-)} = \frac{N_+(K,\uparrow)+N_+(K,\downarrow)-N_-(K,\uparrow)-N_-(K,\downarrow)}{\mathcal{P}_+(K,\uparrow)+\mathcal{P}_+(K,\downarrow)} = \left(\frac{|C_{maj}|^2-|C_{min}|^2}{|C_{maj}|^2+|C_{min}|^2}\right)^2 \quad (S11)$$

By comparing with Eq. (4) and (5) in the main text, we find the difference between $\psi_v^{(n=2)}(K,\uparrow)$ and $\psi_v^{(n=2)}(-K,\uparrow)$, or $\psi_v^{(n=2)}(K,\downarrow)$ and $\psi_v^{(n=2)}(-K,\downarrow)$ is just the permutation of the layer label α and β. Therefore, when considering that the resonant excitation promotes electrons at both K and –K valley, the resulting circular polarization $\rho$ shown in Eq. (S11) will not change. Similarly, we can get exactly opposite $\rho$ by using $\sigma_-$ resonant excitation (see Fig. S1).

### *B. Three kinds of other relaxation channels that could mix LH and RH components.*

The time-dependent relaxation mechanism in bilayer *MX₂* materials could be complicated. Here we consider only three kinds of relaxation channels between the four-degenerate V1 states at K and –K valley that could mix LH and RH ratio and thus affect the circular polarization. Taking $\psi_v^{(n=2)}(K,\uparrow)$ as an example [Fig. 2(a)], after $\sigma_+$ excitation some $\chi_+^{(\alpha)}\otimes|\uparrow\rangle$ electrons have been excited. (i) Intervalley relaxation involving spin flip on the same sector α between $\chi_-^{(\alpha)}\otimes|\downarrow\rangle$ [from $\psi_v^{(n=2)}(-K,\downarrow)$] and $\chi_+^{(\alpha)}\otimes|\uparrow\rangle$. This relaxation has the same mechanism with that of monolayer, which has been previously investigated [2-4]. Such intervalley relaxation naturally involves spin flip, requiring not only large phonon with the momentum of the size of Brillion zone but also magnetic scattering, and thus has long lifetime. The experimental observation in monolayer MoS₂ of $\rho = 1.00 \pm 0.05$ yields a hole valley-spin lifetime of more than 1 ns, which is much longer than the exciton lifetime (~50 ps) [1], indicating that this relaxation channel is heavily suppressed. (ii) Intravalley relaxation involving spin flip on different sectors between $\chi_-^{(\beta)}\otimes|\downarrow\rangle$ [from $\psi_v^{(n=2)}(K,\downarrow)$] and $\chi_+^{(\alpha)}\otimes|\uparrow\rangle$. This process requires interlayer tunneling as well as spin flip, and is thus strongly blocked. Since the spin lifetime is much longer than the exciton lifetime, we ignore the very small spin-flip possibility within the exciton lifetime for simplicity. (iii) Intervalley relaxation with spin conserving on the same sector between $\chi_-^{(\alpha)} \otimes |\uparrow\rangle$ [from $\psi_v^{(n=2)}(-K,\uparrow)$] and $\chi_+^{(\alpha)} \otimes |\uparrow\rangle$. This relaxation mechanism could be efficient since only phonon compensation between different momenta is needed. However, the two states involved $\psi_v^{(n=2)}(K,\uparrow)$ and $\psi_v^{(n=2)}(-K,\uparrow)$ have same LH/RH ratio so that this channel will not affect $\rho$ comparing with the intrastate relaxation within one state even if we assume the relaxation time much faster than exciton recombination.

## *C. Possible reasons for the deviation between theory and experiment.*

The experimentally observed polarized luminescence $\rho_{exp}$ of *MX*$_2$ bilayers is smaller than our calculated theoretical limit (e.g., 14% for WS$_2$ [5]). One possibility that induce the deviation between the two values can be attributed to the unexpected loss during the photoluminescence [6]

$$\frac{\rho_{exp}}{\rho} = \frac{(1-\delta)^2}{1+\tau/\tau_v} \qquad (S12)$$

where $\delta$ denotes the loss of initial excitation by intervalley generation that mainly arises from defects/substrate- and phonon-assisted excitation, and thus could be minimized by high-quality samples and lower temperature. $\tau$ is the exciton lifetime that is determined by both radiative and non-radiative recombination process. It was estimated to be of the order of 10 ps (bilayer WS$_2$), even shorter than that in monolayer [5]. $\tau_v$ denotes the valley lifetime, or orbital lifetime that contains all kinds of relaxation channel between LH and RH components. Besides the first-order relaxation channels discussed above, other relaxation mechanisms including Elliot-Yaffet, D'yakonov-Perel and Bir-Aronov-Pikus in monolayer and bilayer *MX*$_2$ are also compared in recent references [1, 5].

## *D. Better CP performance in MoSe$_2$/BN heterostructure.*

Many band-engineering approaches have been applied to multilayer (*MX*$_2$)$_n$ to transform their indirect band gap into direct [7, 8]. Taking *n* = 2 as an example, when increasing the interlayer distance, the band gap of bilayer *MX*$_2$ steadily becomes direct due to the weaker S-S interaction between two monolayers. Figure 3 shows that after the interlayer distance reaches 4 Å, both VBM and CBM of bilayer *MX*$_2$ relocate at K valley. Therefore, we can insert some inert layers such as BN into bilayer *MX*$_2$ to make the heterostructure ideal platform to realize the intrinsic circular polarization in a centrosymmetric system by spontaneously achieving two factors: 1) Direct band gap engineering to make excitons stable at K valley and thus sufficient direct photoluminescence. 2) Make residual spin on each layer larger (nearly 1) to maximize the theoretical limit of $\rho$.

We next design a heterostructure by intercalating bilayer BN as inert medium into bilayer $MoSe_2$. The latter compound has the lowest $\rho$ for a fixed interlayer distance, as shown in Fig. 3; and the ratio of lattice constant between $MoSe_2$ and BN is 4:3 with the mismatch less than 1%. Figure S2 shows our calculated band structure of centrosymmetric $MoSe_2$/BN/BN/$MoSe_2$ heterostructure with a supercell containing $3\times3$ $MoSe_2$ and $4\times4$ BN within perfect planar alignment, and that of monolayer $MoSe_2$ with $3\times3$ periodicity for comparison. After structure optimization with the van der Waals correction considered by a dispersion-corrected PBE-D2 method [9], we find that BN layer and the adjacent S layer are still flat with the interlayer distance 3.56 Å, indicating weak coupling between the two kinds of 2D materials. The band structure of the heterostructure is nearly the same as that of monolayer $MoSe_2$ in $3\times3$ supercell, except some BN character far from the Fermi level. By analyzing the wavefunction of the band edges we find that the heterostructure performs as two non-interacting $MoSe_2$ monolayers, with each layer carrying opposite local spin polarization $S \approx \pm 1$. In addition, such heterostructure could also avoid the indirect-bandgap problem of bilayer $MoSe_2$ and thus make excitons stable at K valley to reduce the loss of $\rho$. Similar heterostructures have been synthesized by mechanically stacking graphene, h-BN and $MoS_2$ in order on the substrate [10]. We believe that such sandwiched structures as the proposed $MoSe_2$/BN heterostructure, having both optimized polarization anisotropy $\rho \approx 1$ and direct bandgap, are good platforms to realize intrinsic CP in a centrosymmetric system by the current synthesis technology [11].

**Reference**

[1] K. F. Mak, K. He, J. Shan, and T. F. Heinz, Nat. Nanotech. **7**, 494 (2012).

[2] H. Ochoa and R. Roldán, Phys. Rev. B **87**, 245421 (2013).
[3] H.-Z. Lu, W. Yao, D. Xiao, and S.-Q. Shen, Phys. Rev. Lett. **110**, 016806 (2013).
[4] D. Lagarde, L. Bouet, X. Marie, C. R. Zhu, B. L. Liu, T. Amand, P. H. Tan, and B. Urbaszek, Phys. Rev. Lett. **112**, 047401 (2014).
[5] H. Z. Bairen Zhu, Junfeng Dai, Zhirui Gong, Xiaodong Cui, Proc. Natl. Acad. Sci **111**, 11606 (2014).
[6] T. Cao *et al.*, Nat. Commun. **3**, 887 (2012).
[7] P. Koskinen, I. Fampiou, and A. Ramasubramaniam, Phys. Rev. Lett. **112**, 186802 (2014).
[8] P. Johari and V. B. Shenoy, ACS Nano **6**, 5449 (2012).
[9] J. Klimeš, D. R. Bowler, and A. Michaelides, Phys. Rev. B **83**, 195131 (2011).
[10] G.-H. Lee *et al.*, ACS Nano **7**, 7931 (2013).
[11] C. L. Heideman, S. Tepfer, Q. Lin, R. Rostek, P. Zschack, M. D. Anderson, I. M. Anderson, and D. C. Johnson, J. Am. Chem. Soc. **135**, 11055 (2013).

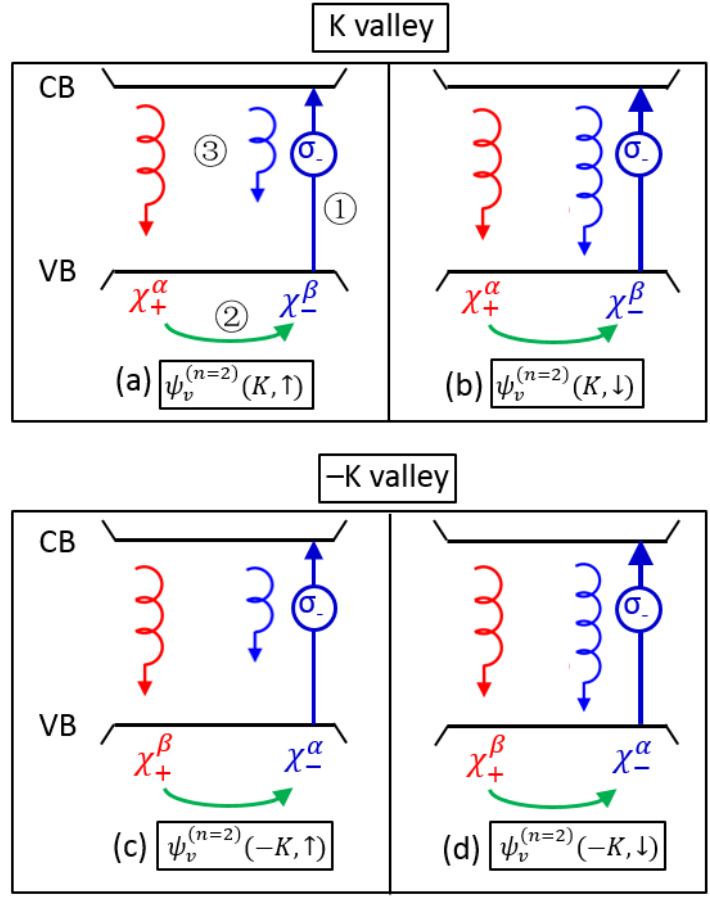

Fig. S1: Schematic diagram of photoluminescence process at K and −K valley in bilayer *MX*$_2$. The notations are same as Fig. 2, but by a resonant $\sigma_-$ excitation.

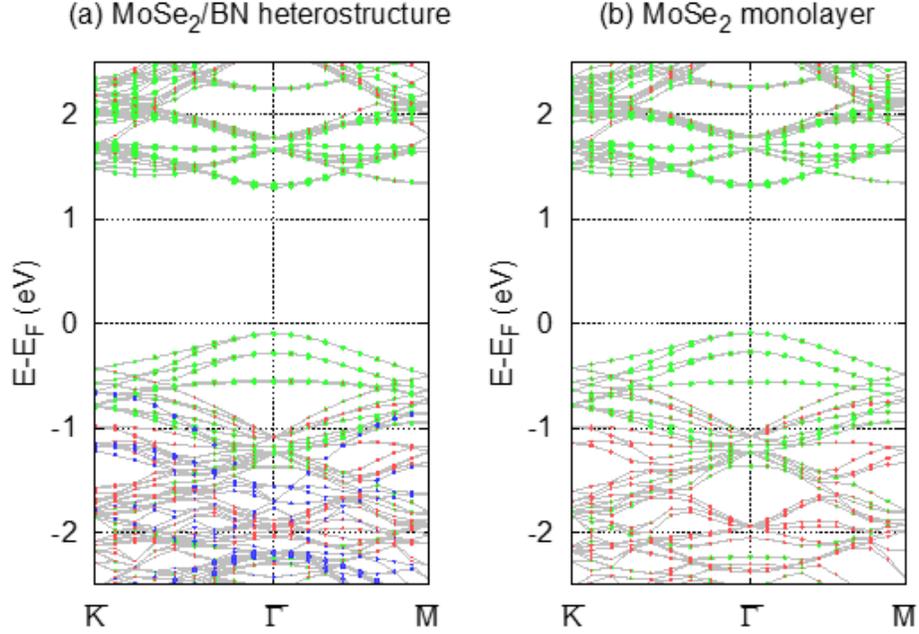

Fig. S2: Band structure with atomic projection of (a) MoSe$_2$/BN/BN/MoSe$_2$ heterostructure and (b) monolayer MoSe$_2$ with 3×3 periodicity. Green, red and blue dots denote the projection of Mo atom, S atom and BN layer, respectively. Note that because of the folding effect of the Brillouin zone, the band edges that locate at K (1/3, 1/3) point in 1×1 unit cell are folded to $\bar{\Gamma}$ in the current 3×3 supercell. By analyzing the components of the wavefunction at the band edges of $\bar{\Gamma}$, we find exactly the same character as those of K in 1×1 unit cell.